\documentclass[journal]{IEEEtran}
\ifCLASSINFOpdf
\else
\fi
\label{begins}

\usepackage{graphicx}
\usepackage{subfig}
\usepackage{cite}
\usepackage{algorithm}
\usepackage{algorithmic}
\usepackage{multicol,multirow,booktabs}
\usepackage{amsmath}
\usepackage{longtable}
\usepackage{rotating}
\usepackage{caption}
\usepackage{amssymb}
\usepackage{pifont}
\usepackage{array}
\usepackage{color}

\begin{document}

%
\title{Cross-Scale Residual Network for Multiple Tasks: Image Super-resolution, Denoising, and Deblocking}
%
%
%

\author{Yuan~Zhou,~\IEEEmembership{Senior~Member,~IEEE}, Xiaoting~Du,
Yeda~Zhang,~\IEEEmembership{Student~Member,~IEEE},
 ~and~Sun-Yuan~Kung,~\IEEEmembership{Life~Fellow,~IEEE}
\thanks{Yuan Zhou, Xiaoting Du and Yeda Zhang are with the School of Electrical and Information Engineering,
Tianjin University, Tianjin 300072, China.}

\thanks{Sun-Yuan Kung, is with the Department of Electrical Engineering,
Princeton University, Princeton, NJ 08540, USA.}}

%
%

\markboth{IEEE Transactions on Cybernetics,~Vol.~XXX, No.~XXX}%
{Shell \MakeLowercase{\textit{\emph{et al.}}}: Bare Demo of IEEEtran.cls for IEEE Journals}
%



\maketitle

\begin{abstract}
\label{sec:0.abstract}
In general, image restoration involves mapping from low quality images to their high-quality counterparts. Such optimal mapping is usually non-linear and learnable by machine learning. Recently, deep convolutional neural networks have proven promising for such learning processing. It is desirable for an image processing network to support well with three vital tasks, namely, super-resolution, denoising, and deblocking. It is commonly recognized that these tasks have strong correlations. Therefore, it is imperative to harness the inter-task correlations. To this end, we propose the cross-scale residual network to exploit scale-related features and the inter-task correlations among the three tasks. The proposed network can extract multiple spatial scale features and establish multiple temporal feature reusage.   Our experiments show that the proposed approach outperforms state-of-the-art methods in both quantitative and qualitative evaluations for multiple image restoration tasks.
\end{abstract}


%
\IEEEpeerreviewmaketitle

\section{Introduction}
\label{sec:1.Introduction}
%
%
%
%
\IEEEPARstart{I}{mage} restoration\cite{han2013enhanced} has been a long-standing problem given its practical value for a variety of low-level vision applications, such as face restoration\cite{chen2019sequential}, semantic segmentation\cite{lin2018scn,nie20183} and target tracking\cite{liu2018robust,zhang2018adaptive}. In general,  image restoration aims to recover clean image $y$ from its corrupted observation $x =H(Y)+v$, where $Y$ is a ground-truth high-quality version of $y$, $H$ is a degradation function, $v$ is additive noise. By accommodating different types of degradation function, the resulting mathematical models target at specific image restoration tasks, such as image super-resolution, denoising, and deblocking.  Image super-resolution reconstructs a high-resolution (HR) image from the low-resolution (LR) counterpart with $H$ being a composite operator of blurring and down-sampling. Image denoising retrieves a clean image from a noisy observation, with $H$ commonly being the identity function and $v$ being additive white Gaussian noise with standard deviation $\sigma $. JPEG image deblocking aims to remove the blocking artifact from a lossy image caused by $H$ corresponding to the JPEG compression function.\par

For decades, model-based optimization and dictionary learning have been dominant in single-task image restoration\cite{yang2010image,buades2008nonlocal,fang2018text,timofte2016seven,jiang2019ensemble}. The recent development of deep learning, especially convolutional neural networks (CNNs), has notably increased progress of image restoration\cite{Zhang2016Beyond, Tai2017MemNet, zhang2017learning,DBLP:journals/corr/abs-1812-10477,Liu2018Multi}. Deep CNNs that enlarge the receptive field or enhance feature reusing provide state-of-the-art results in single-task image restoration, such as single image super-resolution\cite{Kim2016Accurate, Lim2017Enhanced, Haris2018Deep}, image denoising\cite{Harmeling2012Image,Liu2018Multi} or JPEG image artifacts removal\cite{guo2017one,maleki2018blockcnn}, through residual learning and dense connections.\par

It is desirable for an image restoration network to well support all the three aforementioned tasks. Unfortunately, most existing models only perform well in one of these tasks. It is commonly recognized that these tasks happen to have strong correlations. In order to support all the tasks,  the neural network of image restoration must fully harness the inter-task correlations. \par

Moreover,  there exist critical differences on how to best treatment the three tasks. In particular, selection of feature scales is known to significantly impact  the performance on these tasks.  It is also well-known that each of these tasks has its own favorable scales of feature extraction. That is why we propose the cross-scale  residual network (CSRnet) to improve multiple-scales features utilization and the performance on multiple tasks.\par

\begin{figure*}[hbt]
\label{fig:1 CDRN}
 \centering
\includegraphics[scale=.75]{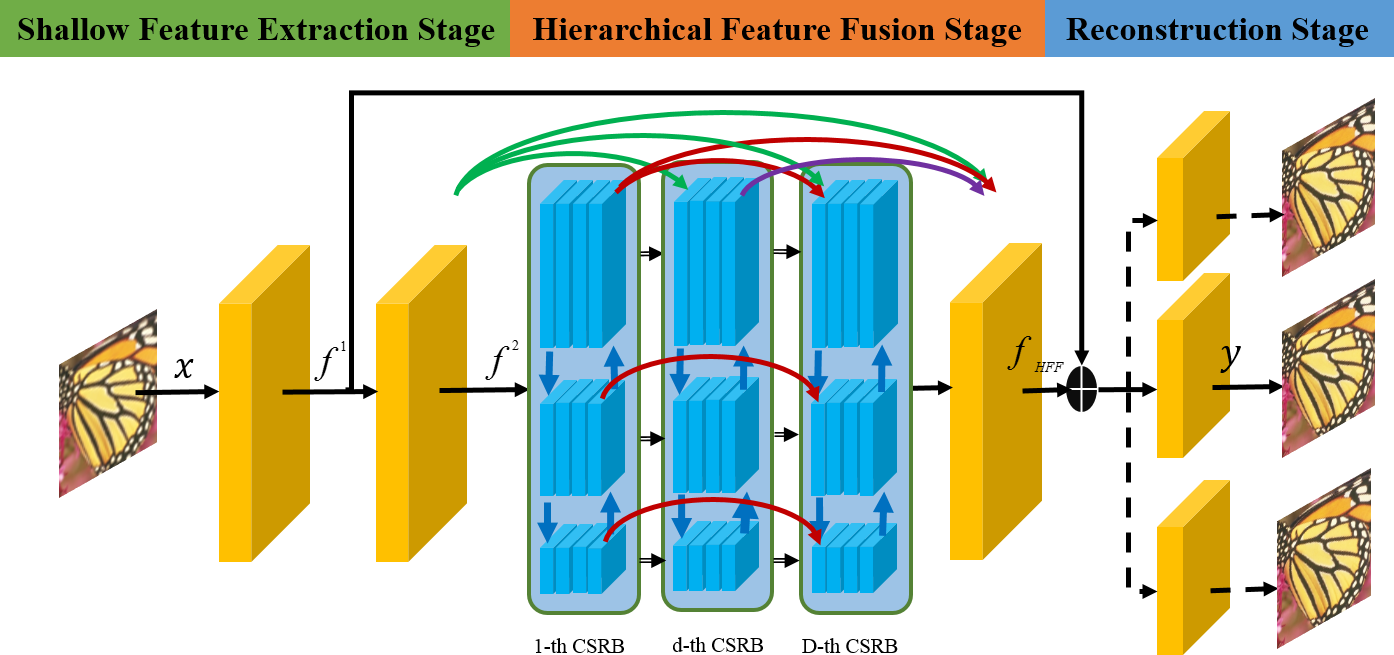}
\caption{Diagram of proposed CSRnet that comprises three parts: shallow feature extraction stage, hierarchical feature fusion stage, and reconstruction stage.}
\end{figure*}

Fig. 1 shows the diagram of the proposed CSRnet, which extracts various features at different scales and fully uses all the hierarchical features throughout the network. Specifically, we propose cross-scale residual blocks (CSRBs) (see Fig.2), whose three states operate at different spatial resolutions, as the building blocks for the CSRnet. The states capture information at different scales, and the intra-block cross-scale connection of each CSRB produces an information flow from the fine to the coarse scale or vice versa. In addition, the inter-block connection combines  information at a given resolution from all the preceding CSRBs,  to provide rich features for the current CSRB. Extensive experimental results verify that each proposed component improves the network performance, and hence the CSRnet outperforms state-of-the-art methods in image super-resolution, denoising, and deblocking.\par

Our main contributions can be summarized in the following aspects:
\begin{enumerate}
  \item
We propose a cross-scale residual network (CSRnet) which simultaneously implements multi-temporal feature reusing and multi-spatial scale feature learning for multiple image restoration tasks, namely, image super-resolution, denoising, and deblocking.
  \item
Cross-scale residual blocks (CSRB) are proposed as the basic building block of the proposed network. The CSRB adaptively learns feature information from different scales. Though deep-learning-based methods have achieved a notable improvement over traditional methods in image restoration domain, most of them learn features from the image space at a single scale, thus cannot handle the scenario of multiple tasks. To this end, we design the CSRB to efficiently extract and adaptively fuse features from different scales for multiple tasks of image restoration.
  \item
  To enhance feature reusing in the blocks and gradient flow during training, we propose two kinds of connections, namely, intra-block cross-scale connection and inter-block connection. The former produces an information flow from the fine to the coarse scale or vice versa. The latter allows the information from preceding blocks to be reused for learning of succeeding block features.
\end{enumerate}\par

\section{Related Work}
\label{sec:2.Related Work}

\subsection{Image super-resolution}

Methods based on CNNs have recently revolutionized the field of image super-resolution. The most commonly used approach is to consider the interpolated low-resolution image as input to the network. Dong et al. \cite{Dong2016Image}  first introduced an end-to-end CNN model called SRCNN to reconstruct interpolated low-resolution images into their high-resolution counterparts. Improvements to the SRCNN include a very deep network for super-resolution (VDSR), which increases the network depth with a smaller filter size and residual learning \cite{Kim2016Accurate}, and a deeply recursive convolutional network (DRCN), which uses recursive layers and multi-supervision  \cite{Kim2015Deeply}. Deep CNN models using block structures \cite{Tai2017Image, Tai2017MemNet}  based on residual units use features from different temporal levels for reconstruction. Although these methods \cite{Kim2016Accurate, zeng2017coupled,Tai2017MemNet, Dong2016Image, Kim2015Deeply, Tai2017Image} have considerably improved super-resolution accuracy, the interpolated low-resolution inputs increase the computational complexity and might introduce additional noise.\par

Given the specificity of image super-resolution, another effective approach directly takes the low-resolution image as input to the CNN\cite{Lim2017Enhanced, ledig2017photo, Tong2017Image, shi2016real, cheon2018generative} for decreasing computational cost. Shi et al.\cite{shi2016real} proposed a sub-pixel convolutional layer to effectively up-sample the low-resolution feature maps in an approach, which is also used in enhanced deep residual networks for super-resolution\cite{Lim2017Enhanced}. Based on dense connection, the SRDenseNet\cite{Tong2017Image} and residual dense network employ dense blocks or residual dense blocks to learn high-level features, whose outputs are concatenated into a final output. In addition, generative adversarial networks have been used for image super-resolution\cite{ledig2017photo, cheon2018generative} to learn adversarial and perceptual content losses that can improve visual quality.\par

\subsection{Image denoising}

Traditional methods such as the BM3D algorithm\cite{Kostadin2007Image} and those based on dictionary learning \cite{Priyam2009Clustering} have improved the performance of image denoising to some extent. Still, methods based on CNNs are more suitable for this task. Xie et al.\cite{Xie2012Image} combined sparse coding with an auto-encoder structure for image denoising. Inspired by residual learning and batch normalization, Zhang et al.\cite{Zhang2016Beyond} proposed the DnCNN model to improve the outcome of image denoising. Mao et al.\cite{Xiao2016Image} proposed a very deep convolutional auto-encoder network (RED) using symmetric skip connections for image denoising and super-resolution.  Du et al.\cite{du2016stacked} proposed
stacked convolutional denoising auto-encoders to map images to hierarchical representations without any label information. Zhang et al. \cite{zhang2017learning} integrated CNN denoisers into model-based optimization for image super-resolution and denoising. Tai et al.\cite{Tai2017MemNet} proposed a very deep persistent memory network (MemNet) that introduces a memory block consisting of a recursive unit and a gate unit to simultaneously perform several image restoration tasks.\par

\subsection{JPEG image deblocking}

Given that JPEG compression often induces severe blocking artifacts and undermines visual quality, image deblocking is particularly important in restoration domain. Chen et al.\cite{Chen2015Trainable} proposed a flexible learning framework based on nonlinear reaction diffusion models for JPEG image deblocking, super-resolution, and denoising. Wang et al.\cite{Zhangyang2016D3} designed a deep dual-domain-based fast restoration model for JPEG image deblocking, which combines prior knowledge from the JPEG compression scheme and the sparsity-based dual-domain approach. Unlike these traditional methods \cite{Chen2015Trainable, Zhangyang2016D3}, JPEG image deblocking based on CNNs is more effective to remove the blocking artifact and improve visual quality. Dong et al.\cite{dong2015compression} proposed an artifact reduction CNN (ARCNN) for JPEG image deblocking. From the method in \cite{Zhangyang2016D3}, the dual-domain CNN proposed by Guo et al. \cite{Guo2016Building} performs joint learning of the discrete cosine transform and pixel domains. To improve visual quality and artistic appreciation, Guo et al.\cite{guo2017one} proposed a one-to-many network for JPEG image deblocking, which measures the output quality using perceptual, naturalness, and JPEG losses.\par

 \subsection{Multiple task image processing}
There exists only a few methods for multi-task image processing. The method proposed by Zhang et al.\cite{zhang2017learning} uses CNN-based denoisers into model-based optimization for image denoising and super-resolution. A very deep persistent memory network \cite{Tai2017MemNet} introduces a memory block to explicitly mine persistent memory through adaptive learning for image denoising, super-resolution, and deblocking. Likewise, Y. Zhang et.al \cite{DBLP:journals/corr/abs-1812-10477} proposed a residual dense network to exploit the hierarchical features from all the convolutional layers in three representative image restoration applications. However, these methods learn image mappings at a single scale, and ignore that different tasks may require features from different scales. \par

\section{Proposed CSRnet for Image Restoration}
\subsection{Architecture}

The proposed CSRnet illustrated in Fig. 1 comprises three stages: shallow feature extraction stage, hierarchical feature fusion stage, and reconstruction stage. They are respectively responsible for extracting shallow image features, fusing abundant feature maps, and adding image details. We denote $x$ and $y$ as the input and output of the CSRnet, respectively.\par

\textbf{Shallow Feature Extraction Stage:} We utilize two convolutional layers to extract shallow features from low-quality input images. The first convolutional layer extracts features from the input image, and the second convolutional layer reduces the dimension of the features. Shallow feature extraction stage can be expressed as:

 \label{eq.1}
 \begin{equation}
 f^{1}=F_{SFE-1}(x)
 \end{equation}
and
 \label{eq.2}
 \begin{equation}
 f^{2}=F_{SFE-2}(f^{1}),
 \end{equation}
where $F_{SFE-1}$ denotes the first convolutional operation,with filter size 7$\times $7. Using a large convolutional kernel can produce a large receptive field which takes a large image context into account. $F_{SFE-2}$ denotes the second convolutional operation, with filter size 3$\times $3. $f^{1}$ is further used for residual learning during the reconstruction stage by skip connection and $f^{2}$ is used as the input for the first CSRB.

\textbf{Hierarchical Feature Fusion Stage:} The CSRnet learns hierarchical features from every CSRB that has identical structure in this stage. If $D$ CSRBs are stacked by inter-block connection, hierarchical feature fusion stage makes full use of the scale state $s= 0$  from each CSRB by:

\label{eq.3}
 \begin{equation}
 f^{HFF}=F_{1\times {1}}([f^{2},f^{output}_{1,0},...,f^{output}_{d,0},f^{output}_{D,0}]),
 \end{equation}
where $[f^{2},f^{output}_{1,0},...,f^{output}_{d,0},f^{output}_{D,0}]$ denotes the concatenation features of the outputs of scale $s=0$  from all the CSRBs and the shallow output from the previous stage, and $F_{1\times {1}}$  introduces a  $1\times {1} $ convolutional layer to adaptively control the dimension of feature maps before inputting reconstruction stage.

\textbf{Reconstruction Stage: } To further improve information flow and reconstruct image details, this stage contains a skip connection and two convolutional layers and is expressed as:

\label{eq.4}
 \begin{equation}
 y=F_{REC}(f^{HFF}+f^{1}),
 \end{equation}
where the skip connection adds output $f^{HFF}$ of the hierarchical feature fusion stage with shallow features $f^{1}$ from shallow feature extraction stage. And  $F_{REC}$ denotes two convolutional operations with filter sizes $3\times {3} $.\par

Given training set $\left \{ x^{(i)},Y^{(i)} \right \}_{i=1}^{N}$ with $N$ training patches and $Y^{(i)}$ is the ground truth high-quality patch corresponding to the low-quality patch $ x^{(i)}$ , we define the loss function of our model with the parameter $\Theta $ set as below:

\label{eq.5}
 \begin{equation}
L(\Theta)=\frac{1}{N}\sum_{i=1}^{N} \left \| Y^{(i)}-y^{(i)} \right \|
 \end{equation}\par

\begin{figure}[hbt]
\label{fig:2 CDRB}

\includegraphics[scale=.48]{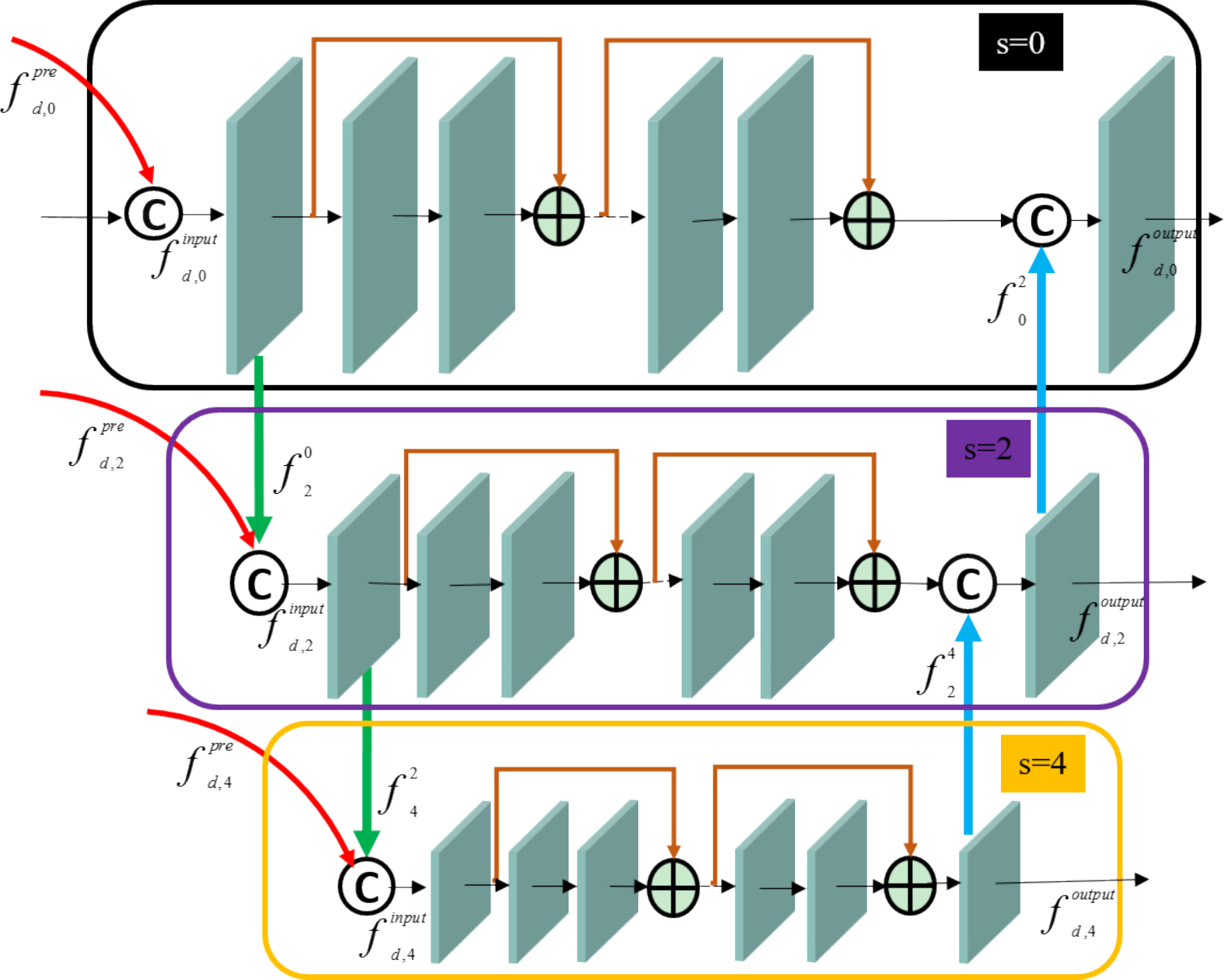}
\caption{ Diagram of proposed CSRB, the key component of CSRnet. }
\end{figure}

\subsection{Cross-Scale Residual Block (CSRB)}

To determine image features at different scales, we propose the CSRB as the key component of the CSRnet. A CSRB adopts three branches using different scales (i.e., $1\times$, $\frac{1}{2}\times$, and $\frac{1}{4}\times$), to enable the use of cross-scale features. It is illustrated in Fig. 2 and detailed as follows.\par

\textbf{Cross-Scale Design: } Unlike models working at a single spatial resolution, the CSRB incorporates information from different scales. Specifically, boxes with different colored edges in Fig. 2 represent the structure designs at different scales.  The boxes with black, purple, and yellow edges indicate scale  $s$ = 0, 2, and 4, respectively. The value of $s$ represents the scale of downsampling, i.e., $1\times$, $\frac{1}{2}\times$, and $\frac{1}{4}\times$. Two colored links, called intra-block cross-scale connections, indicate transitions between the three scales. The green and blue links respectively produce information flow from fine to coarse scale and vice versa. To learn abundant features from the previous blocks, we add the red link, called inter-block connection, at each scale.\par

The input at a given scale ($s$ = 0, 2, 4) in the $d$-th CSRB is computed by concatenating two kinds of features, namely, \textbf{1)} same-scale features ( $f_{d,s}^{pre}$, $s$ = 0, 2, 4) from all the previous CSRBs; \textbf{2)} either shallow features ($f^{2}$, $s$ = 1) or finer-scale feature map ($f_{s}^{s-2}$, $s$ = 2, 4). The finer-scale feature map is obtained by a strided convolution from the higher-resolution layers. The overall inputs of the $d$-th CSRB are given as:

\label{eq.6}
 \begin{equation}
 \begin{split}
f_{d,0}^{input}=F_{1\times {1}}(f_{d,0}^{pre}+f^{2})\\
f_{d,2}^{input}=F_{1\times {1}}(f_{d,2}^{pre}+f_{2}^{0})\\
f_{d,4}^{input}=F_{1\times {1}}(f_{d,4}^{pre}+f_{4}^{2}),
  \end{split}
 \end{equation}
where $F_{1\times {1}}$ denotes a $1\times {1}$ convolutional layer intended to reduce and maintain the dimension for input at different scales, and $f_{s}^{s-2}$ and $f_{d,s}^{pre}$ are described detailedly in the intra-block cross-scale connections and inter-block connections.\par

To facilitate feature specialization at different resolutions, we modify the residual blocks for each scale input of the $d$-th CSRB by removing the batch normalization layers from our network, as performed by Nah et al\cite{Nah2016Deep}. Enhanced deep residual networks for super-resolution\cite{Lim2017Enhanced} have experimentally shown that this simple modification substantially increases performance. The outputs of different scales $s = 0, 2, 4$ in the $d$-th CSRB can be formulated as:

\label{eq.7}
 \begin{equation}
 \begin{split}
f_{d,0}^{output}=F_{1\times{1}}({F_{d,R}^{res}(...(F_{d,1}^{res}(f_{d,0}^{input})))+f_{0}^{2}})\\
f_{d,2}^{output}=F_{1\times{1}}({F_{d,R}^{res}(...(F_{d,1}^{res}(f_{d,2}^{input})))+f_{2}^{4}})\\
f_{d,4}^{output}=F_{1\times {1}}({F_{d,R}^{res}(...(F_{d,1}^{res}(f_{d,4}^{input})))}),
  \end{split}
 \end{equation}
where $F_{d,r}^{res}$ denotes the composite operations of the $r$-th residual block in $d$-th CSRB, including two convolutional layers and an activation function (ReLU). $F_{1\times {1}}$ denotes a $1\times {1}$ convolutional layer, that maintains the dimension for the outputs at different scales of CSRB. $f_{s}^{s+2}$, $s$ = 0, 2 is the coarser-scale feature map obtained by deconvolution from lower-resolution layers, as detailed in the intra-block cross-scale connection.

\textbf{Intra-Block Connection: } Features across different scales can provide various types of information for image restoration. Hence, we propose the intra-block cross-scale connection for producing information flow from fine to coarse scale and vice versa.\par

Finer-scale feature map ($f_{s}^{s-2}$  , $s$ = 2, 4) is produced from the higher-resolution layers by:
\label{eq.8}
 \begin{equation}
f_{s}^{s-2}=F_{down}(f_{d,s-2}^{input}), s=2,4,
 \end{equation}
where $F_{down}$  denotes the down-sample convolutional operations, whose corresponding layer uses a stride size of 2 to reduce the size of the feature map by half.

Likewise, coarser-scale feature map ($f_{s}^{s+2}$ , $s$ = 0, 2) is obtained from lower-resolution layers by:

\label{eq.9}
 \begin{equation}
f_{s}^{s+2}=F_{up}(f_{d,s+2}^{output}),s=0,2,
 \end{equation}
where $F_{up}$  denotes the up-sample convolutional operations, whose corresponding layer uses a stride size of 1/2 to double the size of the feature map.\par

\textbf{Inter-Block Connection: } To enhance feature reusing and gradient flow, we perform inter-block connection that utilizes the information at a given resolution from all previous blocks. The input at a particular scale ($s$ = 0,2,4) of $d$-th CSRB can receive the corresponding scale features of all the preceding CSRBs as follows:

 \label{eq.10}
 \begin{equation}
f_{d,s}^{pre}=\left [f_{1,s}^{output},f_{2,s}^{output},..,f_{d-1,s}^{output} \right ],s=0,2,4,
 \end{equation}
where $\left [f_{1,s}^{output},f_{2,s}^{output},..,f_{d-1,s}^{output} \right ]$  represents the concatenation of features retrieved by all the preceding CSRBs at a particular scale. When d=1, $f_{d,s}^{pre}=0 , s=0,2,4$ .

\begin{figure}[h]
\label{fig:3 coverage analysis}
 \centering
\includegraphics[scale=.50]{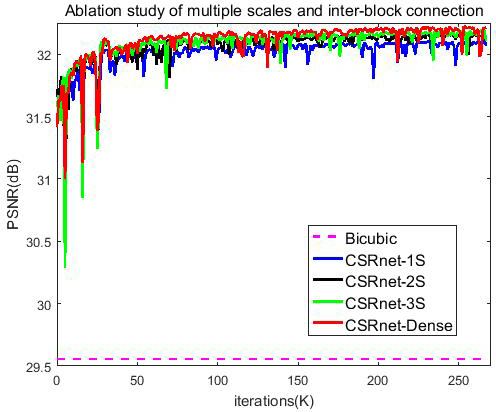}
\caption{Convergence analysis on multiple scales and inter-block connection. The curve for each model is based on the PSNR in 300k iterations of BSD100, with upscaling factor $2 \times$.}
\end{figure}

\section{Experiments}
\label{sec:5 Experiments}

In this section, we first describe the experimental setup including the datasets and network settings of the proposed CSRnet. Then, taking image super-resolution as an example, we evaluate the contributions of different CSRnet components and parameters through an ablation study, and then analyze the effect of the CSRnet depth. Finally, we compare our model with state-of-the-art methods in both objective and subjective aspects on three image restoration tasks, namely, denoising, super-resolution, and deblocking.\par

\subsection{Datasets}
\label{sec:5.1 Training Settings}
%


\begin{figure*}
\captionsetup[subfigure]{labelformat=empty}
\captionsetup{belowskip= -8pt}
    \small
    \subfloat[Ground Truth]{
        \footnotesize
        \includegraphics[width=0.265\linewidth]{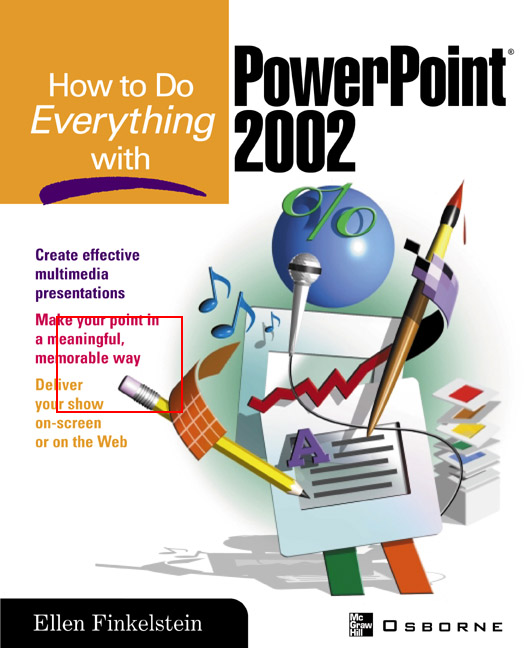}
    }
    \begin{minipage}[b]{0.735\linewidth}
        \subfloat[HR \protect\\(PSNR/SSIM)]{
            \centering
            \includegraphics[width=0.23\linewidth]{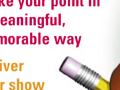}
        }
        \subfloat[Bicubic \protect\\(26.90/0.9434)]{
            \centering
            \includegraphics[width=0.23\linewidth]{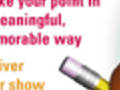}
        }
        \subfloat[SRCNN \protect\\(31.47/0.9790)]{
            \centering
            \includegraphics[width=0.23\linewidth]{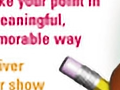}
        }
        \subfloat[VDSR \protect\\(32.76/0.9869)]{
            \centering
            \includegraphics[width=0.23\linewidth]{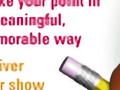}
        }\\
        \subfloat[DRCN\protect\\(32.32/0.9867)]{
            \centering
            \includegraphics[width=0.23\linewidth]{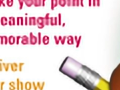}
        }
        \subfloat[LapSRN \protect\\(32.76/0.9878)]{
            \centering
            \includegraphics[width=0.23\linewidth]{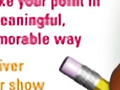}
        }
        \subfloat[MemNet \protect\\ ({\color{blue}{34.46}}/{\color{blue}{0.9902}})]{
            \centering
            \includegraphics[width=0.23\linewidth]{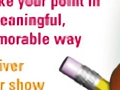}
        }
        \subfloat[Ours \protect\\ ({\color{red}{36.25}}/{\color{red}{0.9931}})]{
            \centering
            \includegraphics[width=0.23\linewidth]{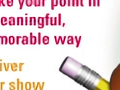}
        }
    \end{minipage}
\caption{Qualitative super-resolution comparison of proposed CSRnet with other models on an image from Set14 dataset with upscaling factor $2\times$.  The CSRnet recovers sharp edges of letters , such as \em{"n"} or \em{"g"}   in the image.}
\label{fig:4}
\end{figure*}

\begin{figure*}
\captionsetup[subfigure]{labelformat=empty}
\captionsetup{belowskip= -8pt}
    \small
    \subfloat[Ground Truth]{
        \footnotesize
        \includegraphics[width=0.245\linewidth]{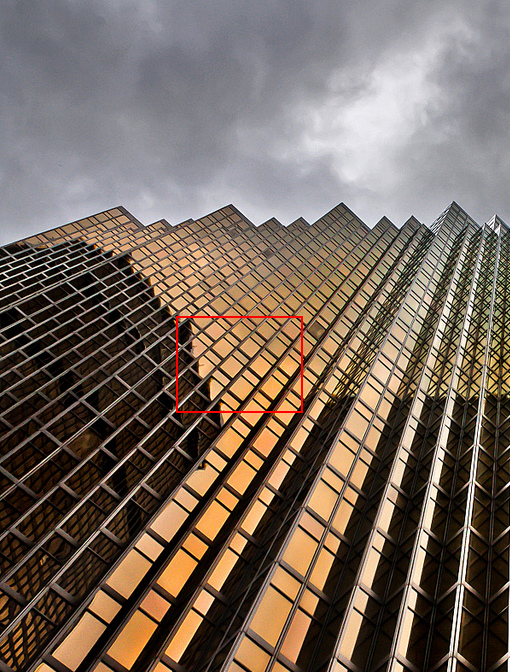}
    }
    \begin{minipage}[b]{0.735\linewidth}
        \subfloat[HR \protect\\(PSNR/SSIM)]{
            \centering
            \includegraphics[width=0.235\linewidth]{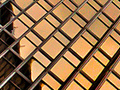}
        }
        \subfloat[Bicubic \protect\\(19.03/0.6517)]{
            \centering
            \includegraphics[width=0.235\linewidth]{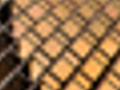}
        }
        \subfloat[SRCNN \protect\\(20.40/0.7402)]{
            \centering
            \includegraphics[width=0.235\linewidth]{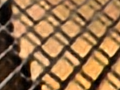}
        }
        \subfloat[VDSR \protect\\(20.82/0.7672)]{
            \centering
            \includegraphics[width=0.235\linewidth]{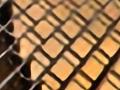}
        }\\
        \subfloat[DRCN\protect\\(20.86/0.7688)]{
            \centering
            \includegraphics[width=0.235\linewidth]{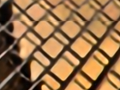}
        }
        \subfloat[LapSRN \protect\\(20.82/0.7699)]{
            \centering
            \includegraphics[width=0.235\linewidth]{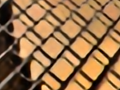}
        }
        \subfloat[MemNet \protect\\ ({\color{blue}{21.45}}/{\color{blue}{0.7886}})]{
            \centering
            \includegraphics[width=0.235\linewidth]{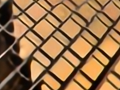}
        }
        \subfloat[Ours \protect\\ ({\color{red}{22.28}}/{\color{red}{0.8121}})]{
            \centering
            \includegraphics[width=0.235\linewidth]{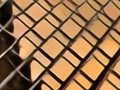}
        }
    \end{minipage}
\caption{Qualitative super-resolution comparison of proposed CSRnet with other models on an image from Urban100 dataset with upscaling factor $4 \times$. Only the CSRnet clearly recovers parallel line structures.}
\label{fig:5}
\end{figure*}

\begin{figure*}
\captionsetup[subfigure]{labelformat=empty}
\captionsetup{belowskip= -8pt}
    \small
    \subfloat[Ground Truth]{
        \footnotesize
        \includegraphics[width=0.220\linewidth]{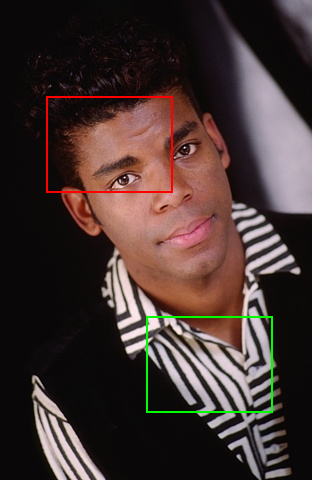}
    }
    \begin{minipage}[b]{0.735\linewidth}
        \subfloat[HR \protect\\(PSNR/SSIM)]{
            \centering
            \includegraphics[width=0.245\linewidth]{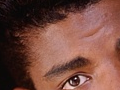}
            \includegraphics[width=0.245\linewidth]{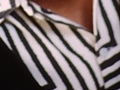}
        }
        \subfloat[VDSR \protect\\(21.35/0.7489)]{
            \centering
            \includegraphics[width=0.245\linewidth]{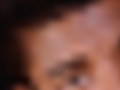}
            \includegraphics[width=0.245\linewidth]{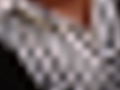}
        }\\
        \subfloat[LapSRN \protect\\(22.40/0.8163)]{
            \centering
            \includegraphics[width=0.245\linewidth]{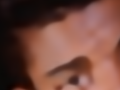}
            \includegraphics[width=0.245\linewidth]{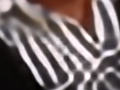}
        }
        \subfloat[Ours \protect\\ ({\color{red}{24.49}}/{\color{red}{0.8498}})]{
            \centering
            \includegraphics[width=0.245\linewidth]{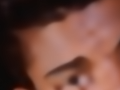}
            \includegraphics[width=0.245\linewidth]{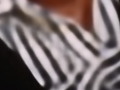}
        }
    \end{minipage}
\caption{Qualitative super-resolution comparison of proposed CSRnet with other models on an image from BSD100 dataset with upscaling factor $8 \times$.Our methods can more realistically restore the man's eyes and his shirt's stripes.}
\label{fig:6}
\end{figure*}

For image super-resolution, we generated the bicubic up-sampled image patches by using function \emph{imresize} in MATLAB \cite{Corke2002A} with option \emph{bicubic} as the input to CSRnet. Following \cite{Dong2016Image, Kim2015Deeply}, we evaluate the proposed model on four popular benchmark datasets, namely Set5 \cite{Bevilacqua2012Neighbor}, Set14 \cite{Zeyde2012On}, BSD100\cite{martin2001database} and Urban100 \cite{Huang2015Single}, with upscaling factors $Up$ = 2$\times$, 4$\times$, and 8$\times$.\par
For image denoising,  we generated noisy patches as CSRnet input by adding Gaussian noise at two levels  $\sigma$ =15, 30 and 50 to the clean patches. Four popular benchmarks, a dataset with 14 common images\cite{Tai2017MemNet}, BSD68\cite{martin2001database}, Urban100 \cite{Huang2015Single} and the BSD testing set with 200 images \cite{Jianchao2010Image}, were used for evaluation.\par

For JPEG image deblocking, we compressed the images using the MATLAB JPEG encoder with compression quality settings $Q$ = 10, 20 as JPEG deblocking input to the CSRnet. Like in \cite{dong2015compression}, we evaluated the CSRnet and comparison methods on the Classic5 and LIVE1 datasets.\par

\subsection{Network Settings}
The objective functions given by Eqn. 5  was optimized via minibatch stochastic gradient descent with backpropagation \cite{dong2015compression}.  To improve the tradeoff between the size of input patches and available computing power, we set the minibatch size to 10, momentum to 0.9, and weight decay to $10^{-4}$.

We use TensorFlow \cite{Abadi2016TensorFlow} to implement the basic CSRnet network. Each convolutional layer, except for the first and final layers, has 32 filters. The first convolution layer has 64 filters, which are used to extract more shallow information. The final convolutional layer has a single feature channel (1 filter), which is used to output the high-quality image. Training the basic CSRnet for image super-resolution roughly required three days on a single GTX 1080 GP (Nvidia Co., Santa Clara, CA, USA). Due to space constraint, we focus on image super-resolution in Sec. IV.C and IV.D, while all three tasks in Sec. IV.E.\par

We evaluate the results for image restoration tasks in terms of the peak signal-to-noise ratio (PSNR) and structural similarity image measurement (SSIM) on the Y channel (luminance) in the YCbCr image space. The other two chrominance channels were directly transformed from the interpolated LR images for displaying the results.
\subsection{Ablation Study}
\label{sec:5.2 Ablation Study}

TABLE I lists the PSNR obtained from the ablation study on the effects of multiple scales and inter-block connection. The baseline (denoted as CSRnet-1S) is at a single scale ($s=0$). To further verify the effectiveness of the multiple spatial scale, CSRnet-2S adds a coarse scale ($s=2$) to baseline CSRnet-1S, and CSRnet-3S adds another coarse scale ($s=4$) to CSRnet-2S. These networks can exchange features among different scales via intra-block cross-scale connections. Among the three networks, CSRnet-3S achieves the best performance on the four testing datasets. To some extent, adding more scales enables a better learning of features, thereby further improves the network performance.\par
\begin{table}[hbt]
\label{tab:1. Ablation study}
\begin{center}
\caption{ PSNR  at upscaling factor $2\times$ obtained from ablation study to evaluate multiple scales and inter-block connection on different datasets. The red entries indicate the best performance.}
{\begin{tabular}{|c|c|c|c|c|}
\hline

\multicolumn{1}{|c|}{Dataset} &\multicolumn{1}{c|}{CSRnet-1S} &\multicolumn{1}{c|}{CSRnet-2S} &
\multicolumn{1}{c|}{CSRnet-3S} &\multicolumn{1}{c|}{CSRnet-Dense} \\
\hline
\multicolumn{1}{|c|}{Set5} &37.853	&37.867	&\color{blue}{37.890}	 &\color{red}{37.999}\\
\hline
\multicolumn{1}{|c|}{Set14} &33.446	&33.500	&\color{blue}{33.543}	 &\color{red}{33.696}\\
\hline
\multicolumn{1}{|c|}{BSD100} &32.115	 &32.155	 &\color{blue}{32.202}	 &\color{red}{32.251} \\
\hline
\multicolumn{1}{|c|}{Urban100} &31.734	&31.890	&\color{blue}{32.075}	 &\color{red}{32.326} \\
\hline

\end{tabular}}{}
\end{center}
\end{table}

\begin{table*}[hbt]
\label{tab:2. results}
\begin{center}
\caption{Average PSNR(dB) / SSIM results of the competing methods for image super-resolution task with upscaling factors  $Up$= $2\times$, $4\times$, and $8\times$ on datasets Set5, Set14,BSD100 and Urban100. The red entries indicate the best performance.}
{
\begin{tabular}{|c|c|c|c|c|c|c|c|c|c|}

\hline
\multirow{2}{*}{Upscaling factor}&\multirow{2}{*}{Methods}&\multicolumn{2}{c|}{Set5}&\multicolumn{2}{c|}{Set14}&\multicolumn{2}{c|}{BSD100}&\multicolumn{2}{c|}{Urban100}
\\
\cline{3-10}
~ & ~ & PSNR & SSIM & PSNR & SSIM & PSNR & SSIM & PSNR & SSIM \\
\hline

\specialrule{0em}{1.5pt}{1.5pt}

\hline
\multirow{10}{*}{$2\times $}
  &Bicubic &33.68  &0.9304 &30.24 &0.8691 &29.56 &0.8440 &26.88	&0.8410\\
~ &SRCNN16\cite{Dong2016Image}	&36.65	&0.9536	&32.45	&0.9067	&31.36	&0.8879	 &29.52	 &0.8965\\
~ &VDSR16\cite{Kim2016Accurate}	&37.53	&0.9587	&33.05	&0.9127	&31.90	&0.8960	 &30.77	 &0.9141\\
~ &DRCN15\cite{Kim2015Deeply}	&37.63	&0.9588	&33.06	&0.9121	&31.85	&0.8942	 &30.76	 &0.9133\\
~ &ESPCN16\cite{shi2016real}	&37.00 &0.9559  &32.75  &0.9098  &31.51 &0.8939  &29.87 &0.9065\\
~ &LapSRN17\cite{Lai2017Deep}	&37.52	&0.9591	&32.99	&0.9124	&31.80	&0.8949	 &30.41	 &0.9101\\

~ &MemNet17\cite{Tai2017MemNet}	&37.78	&0.9597	&33.28	 &0.9142	 &32.08	 &0.8984	 &31.31	&09195\\
~ &WaveResNet17\cite{bae2017beyond}	&37.57	&0.9586	&33.09	 &0.9129	 &32.15	 &0.8995	 &30.96	&0.9169\\
~ &DSRN18\cite{Han2018Image} &37.66  &0.9594  & 33.15  &0.9132  &32.10  &0.8979 & 30.97  &0.9163\\
~ &DRFN18\cite{yang2018drfn} &37.71	&0.9595	 &33.29	 &0.9142	 &32.02	 &0.8979	 &31.08	 &0.9123\\
~ &EEDS19\cite{wang2019end} &37.78  &0.9609  & 33.21  &0.9151  & 31.95  & 0.8963 & -  &-\\
~ &Ours	&\color{red}{38.00}	&\color{red}{0.9613}	&\color{red}{33.70}	 &\color{red}{0.9198}	 &\color{red}{32.25}	 &\color{red}{0.9005}	&\color{red}{32.33}	 &\color{red}{0.9298}\\
\hline
\specialrule{0em}{2.5pt}{2.5pt}

\hline
\multirow{10}{*}{$4\times $}

  &Bicubic	&28.42	&0.8109	&26.10	&0.7023	&25.96	&0.6678	&23.15	&0.6574\\
~ &SRCNN16\cite{Dong2016Image}	&30.48	&0.8628	&27.50	&0.7513	&26.91	&0.7103	 &24.53	 &0.7226\\
~ &VDSR16\cite{Kim2016Accurate} &31.35	&0.8838	&28.03	&0.7678	&27.29	&0.7252	 &25.18	 &0.7525\\
~ &DRCN15\cite{Kim2015Deeply}	&31.53	&0.8854	&28.04	&0.7673	&27.24	&0.7233	 &25.14	 &0.7511\\
~ &ESPCN16\cite{shi2016real}	&30.66 &0.8646  &27.71 &0.7562  &26.98 &0.7124 &24.60 &0.7360\\
~ &LapSRN17\cite{Lai2017Deep}	&31.54	&0.8866	&28.19	&0.7694	&27.32	&0.7264	 &25.21	 &0.7553\\
~ &WaveResNet17\cite{bae2017beyond}	&31.52& 0.8864 &28.11 &0.7699 &27.32&0.7266 &25.36&0.7614\\
~ &MemNet17\cite{Tai2017MemNet}	&31.74	&0.8893	&28.26	 &0.7723	 &27.40	 &0.7281	&25.50	&0.7630\\
~ &DRFN18\cite{yang2018drfn} &31.55	&0.8861	&28.30	 &0.7737	 &27.39	 &0.7293	 &25.45	 &0.7629\\
~ &DSRN18\cite{Han2018Image} &31.40  &0.8834  &28.07  & 0.7702 &27.25 & 0.7243  &25.08  &0.7471\\
~ &EEDS19\cite{wang2019end} &31.53  &0.8869  & 28.13  &0.7698  & 27.35  & 0.7263 & -  &-\\
~ &Ours	&\color{red}{32.12}	&\color{red}{0.8929}	&\color{red}{28.51}	 &\color{red}{0.7788}	 &\color{red}{27.55}	 &\color{red}{0.7343}	&\color{red}{26.10}	 &\color{red}{0.7842}\\
\hline
\specialrule{0em}{2.5pt}{2.5pt}

\hline
\multirow{10}{*}{$8\times $}
  &Bicubic	&24.40 &0.6045  &23.19 &0.5110  &23.67 &0.4808  &20.74  &0.4841\\
~ &SRCNN16\cite{Dong2016Image}	&25.34  &0.6471 & 23.86  &0.5443  &24.14 &0.5043 &21.29 &0.5133\\
~ &VDSR16\cite{Kim2016Accurate}	&25.73 &0.6743  &23.20 &0.5110  &24.34 &0.5169  &21.48 &0.5289\\
~ &DRCN15\cite{Kim2015Deeply}	&25.93 &0.6743  &24.25 &0.5510  &24.49 &0.5168 &21.71 &0.5289\\
~ &ESPCN16\cite{shi2016real}	&25.75 &0.6738  &24.21 &0.5109  &24.37 &0.5277 &21.59 &0.5420\\
~ &LapSRN17\cite{Lai2017Deep}	&26.15 &0.7028  &24.45 &0.5792  &24.54 &0.5293 &21.81 &0.5555\\
~ &Ours	&\color{red}{26.44}	&\color{red}{0.7523}	&\color{red}{24.65}	 &\color{red}{0.6316}	 &\color{red}{24.76}	 &\color{red}{0.5924}	&\color{red}{22.31}	 &\color{red}{0.6059}\\
\hline

\end{tabular}}{}
\end{center}
\end{table*}

\begin{table*}[hbt]
\label{tab:4. results}
\begin{center}
\caption{Average PSNR(dB)/SSIM results of the competing methods for image denoising task with noise levels $\sigma$ =15, 30 and 50 on datasets S14 and BSD200. The red and blue entries  indicate the best.}
{
\begin{tabular}{|c|c|c|c|c|c|c|c|c|c|c|c|c|c|c|c|}

\hline
\multirow{2}{*}{Dataset}&\multirow{2}{*}{$\sigma$}&\multicolumn{1}{c|}{BM3D07\cite{Kostadin2007Image}}&\multicolumn{1}{c|}{PGPD15\cite{xu2015patch}}&\multicolumn{1}{c|}{TNRD15\cite{Chen2015Trainable}}
&\multicolumn{1}{c|}{DnCNN16\cite{Zhang2016Beyond}}&\multicolumn{1}{c|}{MemNet17\cite{Tai2017MemNet}}&\multicolumn{1}{c|}{FOCNet19\cite{jia2019focnet}}&\multicolumn{1}{c|}{Ours}\\
\cline{3-9}
~ & ~ & PSNR/SSIM & PSNR/SSIM & PSNR/SSIM & PSNR/SSIM & PSNR/SSIM  & PSNR/SSIM & PSNR/SSIM  \\
\hline

\specialrule{0em}{1.5pt}{1.5pt}

\hline
\multirow{2}{*}{14 images}
    &15  &- /-    &32.01/0.8984 &32.23 /0.9041 &32.56/0.9110 &-/- &-/- &\color{red}{32.86}/\color{red}{0.9162} \\
  &30  &28.49/0.8204  &26.19/0.7442 &27.03/0.7305 &29.04/0.8389 &29.22/0.8444 &-/-&\color{red}{29.45}/\color{red}{0.8516} \\

~  &50  &26.08/0.7427  &24.71/0.6913  &26.27/0.7502  &26.66/0.7678 &26.91/0.7775&-/- &\color{red}{27.09}/\color{red}{0.7875}\\

\hline

\specialrule{0em}{1.5pt}{1.5pt}

\hline
\multirow{2}{*}{BSD200}
    &15  &-/-    &31.38/0.8776&31.65/0.8890 &31.99/0.8976 &-/- &-/- &\color{red}{32.16}/\color{red}{0.9017} \\
  &30  &27.31 /0.7755  &27.33 /0.7717 &26.76/0.7101 &28.52/0.8094 &28.04/0.8053 &-/- &\color{red}{28.82}/\color{red}{0.8220} \\

~  &50  &25.06/0.6831  &25.18/0.6841  &26.02/0.7111  &26.31/0.7287 &25.86/0.7202&-/- &\color{red}{26.64}/\color{red}{0.7487}\\

\hline

\specialrule{0em}{1.5pt}{1.5pt}

\hline
\multirow{2}{*}{Urban100}
    &15  &32.34/0.9220    &32.18/0.9154 &31.98/0.9187 &32.67/0.9250 &-/- &33.15/-&\color{red}{33.35}/\color{red}{0.9361} \\
  &30  &-/-  &28.59/0.8495 &26.79/0.7612 &28.88 /0.8566 &29.11/0.8633 &-/-&\color{red}{30.02}/\color{red}{0.8895} \\

~  &50  &25.94/0.7791  &26.00/0.7760  &25.71/0.7756  &26.28/0.7869 &26.64/0.8023&27.40/- &\color{red}{27.56}/\color{red}{0.8373}\\

\hline

\specialrule{0em}{2.5pt}{2.5pt}

\hline
\multirow{2}{*}{BSD68}
    &15  &31.08/0.8722    &31.13/0.8693 &31.42/0.8822 &31.73/0.8906 &-/- &31.83/-&\color{red}{31.87}/\color{red}{0.8952} \\
   &30  &-/-    &27.81/0.7693 &26.76/0.7108 &28.36/0.7999 &28.46/0.8039 &-/-&\color{red}{28.61}/\color{red}{0.8105} \\
~  &50  &25.62 /0.6869 &25.75/0.6869 &25.97/0.7021  &26.23/0.7189 &26.37/0.7290 &26.50/- &\color{red}{26.53}/\color{red}{0.7372} \\

\hline

\end{tabular}}{}
\end{center}
\end{table*}

\begin{table*}[hbt]
\label{tab:5. results}
\begin{center}
\caption{ Average PSNR(dB) / SSIM results of the competing methods for JPEG image deblocking  task with quality factors $Q$ = 10, 20 on datasets Classic5 and LIVE1. The red entries  indicate the best performance.}
{
\begin{tabular}{|c|c|c|c|c|c|c|c|c|c|c|c|c|c|c|c|}

\hline
\multirow{2}{*}{Dataset}&\multirow{2}{*}{Q}&\multicolumn{1}{c|}{JPEG}&\multicolumn{1}{c|}{ARCNN15\cite{dong2015compression}}
&\multicolumn{1}{c|}{TNRD15\cite{Chen2015Trainable}}&\multicolumn{1}{c|}{DnCNN16\cite{Zhang2016Beyond}}&\multicolumn{1}{c|}{MemNet17\cite{Tai2017MemNet}}
&\multicolumn{1}{c|}{IACNN19\cite{kim2019pseudo}}&\multicolumn{1}{c|}{Ours}\\
\cline{3-9}
~ & ~ & PSNR/SSIM & PSNR/SSIM & PSNR /SSIM & PSNR/SSIM & PSNR/SSIM & PSNR /SSIM & PSNR /SSIM  \\
\hline

\specialrule{0em}{1.5pt}{1.5pt}

\hline
\multirow{2}{*}{Classic5}
  &10  &27.82/0.7595 &29.03 /0.7929 &29.28 /0.7992  &29.40/0.8026 &29.69/0.8107 &29.43/0.8070&\color{red}{30.03}/\color{red}{0.8199} \\

~  &20 &30.12/0.8344  &31.15/0.8517  &31.47/0.8576  &31.63/0.8610 &31.90/0.8658 &31.64/0.8628&\color{red}{32.21}/\color{red}{0.8708}\\

\hline
\specialrule{0em}{2.5pt}{2.5pt}

\hline
\multirow{2}{*}{LIVE1}

   &10 &27.77/0.7730 &28.96/0.8076 &29.15/0.8111 &29.19/0.8123 &29.45/0.8193 &29.34/0.8199&\color{red}{29.72}/\color{red}{0.8257} \\
~  &20 &30.07/0.8512 &31.29 /0.8733  &31.46/0.8769  &31.59/0.8802  &31.83/0.8846&31.73/0.8848&\color{red}{32.08}/\color{red}{0.8886} \\

\hline

\end{tabular}}{}
\end{center}
\end{table*}


Then, we add inter-block connections to CSRnet-3S and denote the resulting network as CSRnet-Dense, which corresponds to the complete CSRnet. Compared to the previous CSRnet variants, CSRnet-Dense achieves the best results on the four testing datasets, which verifies the effect of inter-block connection. Through the inter-block connections, each component is able to contribute to information and gradient flow through the network.\par

To demonstrate the convergence of the four evaluated CSRnet variants, we determined PSNR curves shown in Fig. 3 with bicubic results being the reference. The four models have a stable training process without obvious performance degradation. In addition, multiple scales and inter-block connection not only accelerate convergence but also notably improve performance.\par

\subsection{Depth analysis of our network}
\label{sec:5.3 Depth analysis of our network}

\begin{table}[hbt]
\label{tab:2. Comparison on different network depths}
\begin{center}
\caption{PSNR at upscaling factor $2\times$ retrieved from different network depths determined by the number of CSRBs on different datasets.  The red entries indicate the best performance. }

{\begin{tabular}{|c|c|c|c|}
\hline

\multicolumn{1}{|c|}{Dataset} &\multicolumn{1}{c|}{B4R6} &\multicolumn{1}{c|}{B6R6} &
\multicolumn{1}{c|}{B8R6}  \\
\hline
\multicolumn{1}{|c|}{Set5} &37.923 &\color{blue}{37.959}	&\color{red}{37.999}\\
\hline
\multicolumn{1}{|c|}{Set14} &33.635	&\color{blue}{33.678}	&\color{red}{33.696}\\
\hline
\multicolumn{1}{|c|}{BSD100} &32.193  &\color{blue}{32.215}	 &\color{red}{32.251} \\
\hline
\multicolumn{1}{|c|}{Urban100} &31.998 &\color{blue}{32.187}	&\color{red}{32.326} \\
\hline

\end{tabular}}{}
\end{center}
\end{table}
Besides different architectures, we evaluated different depths of the proposed CSRnet. The network depth is related to two basic parameters: number $D$ of CSRBs and number $R$ of residual blocks per CSRB. In this study, we only tested the effect of number of blocks, $D$, by setting up three structures: $D4R6, D6R6, D8R6$. TABLE V lists the PSNR obtained from image super-resolution of these networks on the four evaluated datasets, Set5, Set14, BSD100 and Urban100 with upscaling factor 2$\times$. Increasing the number of CSRBs considerably improves the PSNR in the datasets given the increased network depth, which in turn retrieves more hierarchical features for improving performance.\par

\subsection{Comparisons with State-of-the-Art Models}
\label{sec:5.4 Comparisons with State-of-the-Art Models}

We compared the CSRnet with the state-of-the-art models for three restoration tasks, namely, image super-resolution, image denoising, and JPEG image deblocking.\par

\subsubsection{Image Super-resolution}
\label{sec:5.4.1 Image Super-resolution}

Regarding image super-resolution, we quantitatively compared the proposed CSRnet with eight state-of-the-art methods, namely, namely SRCNN16\cite{Dong2016Image}, VDSR16\cite{Kim2016Accurate},
DRCN15\cite{Kim2015Deeply}, ESPCN16\cite{shi2016real}, LapSRN17\cite{Lai2017Deep}, MemNet17\cite{Tai2017MemNet}, WaveResNet17\cite{bae2017beyond}, DRFN18\cite{yang2018drfn}, DSRN18\cite{Han2018Image} and EEDS19\cite{wang2019end}. For a fair comparison, we evaluated all the methods on the luminance channel for all upscaling factors. The comparison results on the four evaluated datasets for three upscaling factors ($Up$=2$\times$, 4$\times$, 8$\times$) are listed in TABLE II. The proposed CSRnet substantially outperforms the comparison models over the different upscaling factors and test datasets. On the Urban100 dataset, the CSRnet outperforms the second-best method by a PSNR gain of 0.60 dB at upscaling factor 4 $\times $. On the BSD100, the CSRnet achieves a PSNR gain of only 0.15 dB compared with the second-best method. Similar results occur at other scales and with respect to other comparison models. Hence, the proposed CSRnet performs better especially on structured images with similar geometric patterns across various spatial resolutions, such as urban scenes (Urban100). What's more worth mentioning is the SSIM performance of our method at upscaling factors 8 $\times $. The SSIM value of CSRnet can be 0.05$\sim$0.07 higher than LapSRN at upscaling factor 8 $\times $, however, at other upscaling factors, the SSIM value of CSRnet is no more than 0.02 higher than LapSRN. This strongly proves that our method can retain higher structural similarity under larger upscaling factor.\par

Besides the quantitative comparison, Figs. 4, 5 and 6 show visual comparisons among the evaluated methods. Fig. 4 shows that the proposed CSRnet reconstructs clearer letters than the other models on an image from the Set14 dataset at upscaling factor 2 $\times $.  Likewise, Fig. 5 shows that the CSRnet clearly recovers the parallel line structures on an image from the Urban100 dataset at upscaling factor 4 $\times $, whereas the other models retrieve obvious distortions.  In Fig. 6, our model more realistically restores the man's eyes and his shirt's stripes for an image from the BSD100 dataset at upscaling factor 8 $\times $, whereas other methods are highly distorted. Overall, the CSRnet outperforms the other evaluated models both quantitatively and qualitatively.\par

\begin{figure*}
\captionsetup[subfigure]{labelformat=empty}
\captionsetup{belowskip= -8pt}
    \small
    \subfloat[Ground Truth]{
        \footnotesize
        \includegraphics[width=0.242\linewidth]{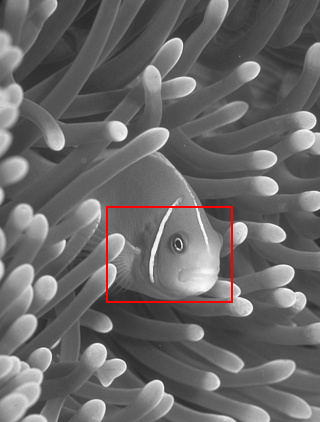}
    }
    \begin{minipage}[b]{0.735\linewidth}
        \subfloat[HR \protect\\(PSNR/SSIM)]{
            \centering
            \includegraphics[width=0.235\linewidth]{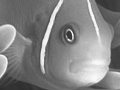}
        }
        \subfloat[Noise \protect\\(24.62/0.4561)]{
            \centering
            \includegraphics[width=0.235\linewidth]{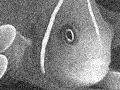}
        }
        \subfloat[IRCNN\protect\\(29.81/0.7096)]{
            \centering
            \includegraphics[width=0.235\linewidth]{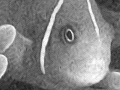}
        }
        \subfloat[BM3D \protect\\(35.29/0.9376)]{
            \centering
            \includegraphics[width=0.235\linewidth]{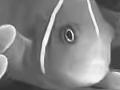}
        }\\
        \subfloat[TRND \protect\\(35.44/0.9362)]{
            \centering
            \includegraphics[width=0.235\linewidth]{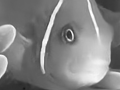}
        }
        \subfloat[PGPD\protect\\(35.09/0.9309)]{
            \centering
            \includegraphics[width=0.235\linewidth]{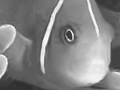}
        }
        \subfloat[DnCNN \protect\\{\color{blue}{36.23}}/{\color{blue}{0.9459}}]{
            \centering
            \includegraphics[width=0.235\linewidth]{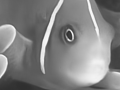}
        }
        \subfloat[Ours \protect\\ ({\color{red}{36.98}}/{\color{red}{0.9559}})]{
            \centering
            \includegraphics[width=0.235\linewidth]{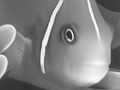}
        }
    \end{minipage}
\caption{Qualitative comparison of our methods with other methods on an image from BSD68 with noise level 15. Our method restores the shape of the coral fish in the water, especially the lip of fish.}
\label{fig:8}
\end{figure*}

\begin{figure*}
\captionsetup[subfigure]{labelformat=empty}
\captionsetup{belowskip= -8pt}
    \small
    \subfloat[Ground Truth]{
        \footnotesize
        \includegraphics[width=0.242\linewidth]{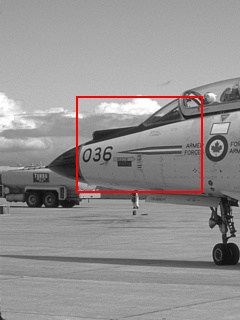}
    }
    \begin{minipage}[b]{0.735\linewidth}
        \subfloat[HR \protect\\(PSNR/SSIM)]{
            \centering
            \includegraphics[width=0.235\linewidth]{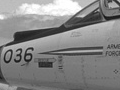}
        }
        \subfloat[Noise \protect\\(18.68/0.2271)]{
            \centering
            \includegraphics[width=0.235\linewidth]{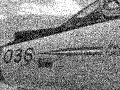}
        }
        \subfloat[IRCNN\protect\\(24.70/0.4889)]{
            \centering
            \includegraphics[width=0.235\linewidth]{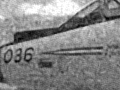}
        }
        \subfloat[TRND \protect\\(27.07/0.7206)]{
            \centering
            \includegraphics[width=0.235\linewidth]{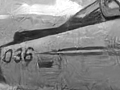}
        }\\
        \subfloat[PGPD\protect\\(28.74/0.8659)]{
            \centering
            \includegraphics[width=0.235\linewidth]{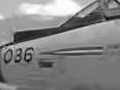}
        }
        \subfloat[DnCNN \protect\\(28.77/0.8774)]{
            \centering
            \includegraphics[width=0.235\linewidth]{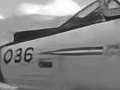}
        }
        \subfloat[MemNet \protect\\ ({\color{blue}{29.08}}/{\color{blue}{0.8977}})]{
            \centering
            \includegraphics[width=0.235\linewidth]{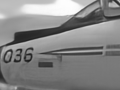}
        }
        \subfloat[Ours \protect\\ ({\color{red}{29.24}}/{\color{red}{0.9052}})]{
            \centering
            \includegraphics[width=0.235\linewidth]{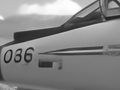}
        }
    \end{minipage}
\caption{Qualitative comparison of our methods with other methods on an image from BSD200 with noise level 30. Our method can recover the clearest identification information on the plane.}
\label{fig:7}
\end{figure*}

\begin{figure*}
\captionsetup[subfigure]{labelformat=empty}
\captionsetup{belowskip= -8pt}
    \small
    \subfloat[Ground Truth]{
        \footnotesize
        \includegraphics[width=0.242\linewidth]{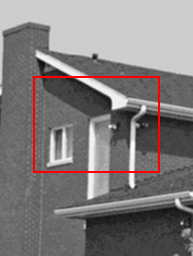}
    }
    \begin{minipage}[b]{0.735\linewidth}
        \subfloat[HR \protect\\(PSNR/SSIM)]{
            \centering
            \includegraphics[width=0.235\linewidth]{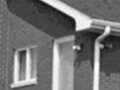}
        }
        \subfloat[Noise \protect\\(14.82/0.1459)]{
            \centering
            \includegraphics[width=0.235\linewidth]{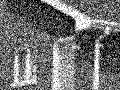}
        }
        \subfloat[IRCNN\protect\\(22.41/0.3699)]{
            \centering
            \includegraphics[width=0.235\linewidth]{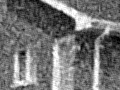}
        }
        \subfloat[TRND \protect\\(29.52/0.8408)]{
            \centering
            \includegraphics[width=0.235\linewidth]{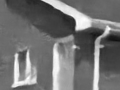}
        }\\
        \subfloat[PGPD\protect\\(30.29/0.8438)]{
            \centering
            \includegraphics[width=0.235\linewidth]{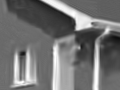}
        }
        \subfloat[DnCNN \protect\\(30.19/0.8501)]{
            \centering
            \includegraphics[width=0.235\linewidth]{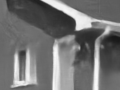}
        }
        \subfloat[MemNet \protect\\ ({\color{blue}{30.90}}/{\color{blue}{0.8659}})]{
            \centering
            \includegraphics[width=0.235\linewidth]{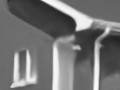}
        }
        \subfloat[Ours \protect\\ ({\color{red}{31.63}}/{\color{red}{0.8733}})]{
            \centering
            \includegraphics[width=0.235\linewidth]{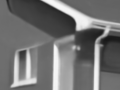}
        }
    \end{minipage}
\caption{Qualitative comparison of our methods with other methods on an image from S14 with noise level 50. Our method restores the window of the house more clearly.}
\label{fig:8}
\end{figure*}

\subsubsection{ Image Denoising}
\label{sec:5.4.2  Image Denoising}

We trained the proposed CSRnet by using the gray images and compared the results to those obtained from eight denoising methods: BM3D07\cite{Kostadin2007Image}, TNRD15 \cite{Chen2015Trainable},PGPD15\cite{xu2015patch} DnCNN16\cite{Zhang2016Beyond}, IRCNN17\cite{zhang2017learning}, RED18\cite{yang2018drfn}, MemNet17\cite{Tai2017MemNet} and FOCNet19\cite{jia2019focnet}. TABLE III lists the average PSNR/SSIM results of the evaluated methods on four benchmark datasets for three noise levels. The PSNR values of CSRnet is better than those of the second-best method at any noise level or any dataset. Like for super-resolution, Figs. 7 , 8 and 9 show visual comparisons among the evaluated methods on an image from BSD68 with the noise level $\sigma$ = 15, an image from BSD200 with the noise level $\sigma$ = 30 and an image from S14 with the noise level $\sigma$ = 50. The proposed CSRnet recovers relatively sharper and clearer images than the other methods, thus being more faithful to the ground truth.\par

\begin{figure*}
\captionsetup[subfigure]{labelformat=empty}
\captionsetup{belowskip= -8pt}
    \small
    \subfloat[Ground Truth]{
        \footnotesize
        \includegraphics[width=0.294\linewidth]{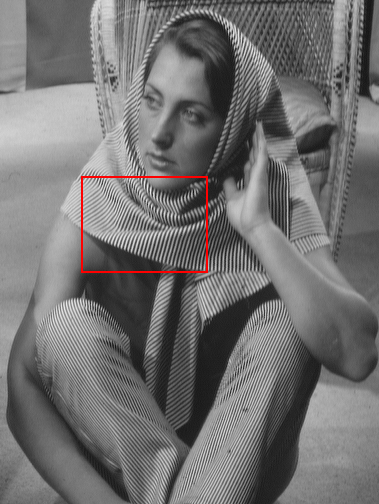}
    }
    \begin{minipage}[b]{0.755\linewidth}
        \subfloat[HR \protect\\(PSNR/SSIM)]{
            \centering
            \includegraphics[width=0.291\linewidth]{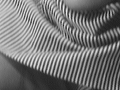}
        }
        \subfloat[Deblocking \protect\\(25.71/0.7610)]{
            \centering
            \includegraphics[width=0.291\linewidth]{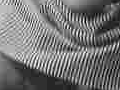}
        }
        \subfloat[ARCNN\protect\\(26.83/0.7951)]{
            \centering
            \includegraphics[width=0.291\linewidth]{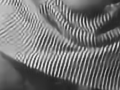}
        }
        \\
        \subfloat[DnCNN\protect\\(27.50/0.8150)]{
            \centering
            \includegraphics[width=0.291\linewidth]{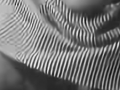}
        }
        \subfloat[MemNet \protect\\ ({\color{blue}{28.04}}/{\color{blue}{0.8335}})]{
            \centering
            \includegraphics[width=0.291\linewidth]{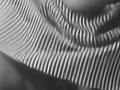}
        }
        \subfloat[Ours \protect\\ ({\color{red}{28.87}}/{\color{red}{0.8542}})]{
            \centering
            \includegraphics[width=0.291\linewidth]{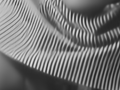}
        }
    \end{minipage}
\caption{Qualitative comparison of our methods with other methods on an image from Classic5 with quality factor 10. Our method recovers lighthouse.}
\label{fig:10}
\end{figure*}

\begin{figure*}
\captionsetup[subfigure]{labelformat=empty}
\captionsetup{belowskip= -8pt}
    \small
    \subfloat[Ground Truth]{
        \footnotesize
        \includegraphics[width=0.302\linewidth]{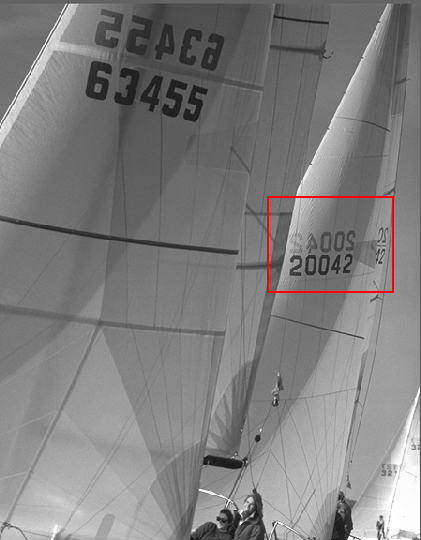}
    }
    \begin{minipage}[b]{0.755\linewidth}
        \subfloat[HR \protect\\(PSNR/SSIM)]{
            \centering
            \includegraphics[width=0.287\linewidth]{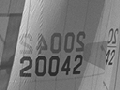}
        }
        \subfloat[Deblocking \protect\\(29.39/0.7998)]{
            \centering
            \includegraphics[width=0.287\linewidth]{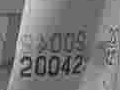}
        }
        \subfloat[ARCNN\protect\\(34.11/0.8981)]{
            \centering
            \includegraphics[width=0.287\linewidth]{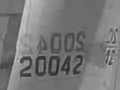}
        }
        \\
        \subfloat[DnCNN\protect\\(34.59/0.9067)]{
            \centering
            \includegraphics[width=0.287\linewidth]{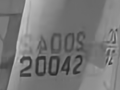}
        }
        \subfloat[MemNet \protect\\ ({\color{blue}{35.06}}/{\color{blue}{0.9136}})]{
            \centering
            \includegraphics[width=0.287\linewidth]{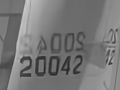}
        }
        \subfloat[Ours \protect\\ ({\color{red}{35.33}}/{\color{red}{0.9196}})]{
            \centering
            \includegraphics[width=0.287\linewidth]{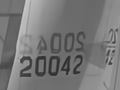}
        }
    \end{minipage}
\caption{Qualitative comparison of our methods with other methods on an image from LIVE1 with quality factor 20. Our method recovers lighthouse.}
\label{fig:10}
\end{figure*}
\subsubsection{ JPEG image Deblocking}
\label{sec:5.4.3  JPEG image Deblocking}

We applied the proposed CSRnet for deblocking considering only on the Y channel and compared it with four existing methods: ARCNN15\cite{dong2015compression}, TNRD15\cite{Chen2015Trainable}, DnCNN16\cite{Zhang2016Beyond}, MemNet17\cite{Tai2017MemNet} and IACNN19\cite{kim2019pseudo}. Table IV lists the average PSNR/SSIM of the evaluated methods on two benchmark datasets, namely, Classic5 and LIVE1, for quality factors of 10 and 20. The CSRnet outperforms IACNN19\cite{kim2019pseudo}, the current state-of-the-art method, by more than 0.60 and 0.57 dB in Classic5 dataset, and 0.38 and 0.35 dB in the LIVE1 dataset with quality factors of 10 and 20 respectively. Fig. 10 and 11 show visual comparisons for JPEG image deblocking. ARCNN, DnCNN, and MemNet were compared using their public codes. Clearly, CSRnet more effectively removes the blocking artifact and restores detailed textures than the comparison methods.\par

\section{Conclusion}
\label{sec:6 conclusion}
 This paper presents the CSRnet, a deep network intended to exploit scale-related features and the inter-task correlations among the three tasks: super-resolution, denoising, and deblocking. Several CSRBs are stacked in the CSRnet and adaptively learn image features at different scales. The same-resolution outputs from all the previous CSRBs are used by the current CSRB via inter-block connections for reusing information. The intra-block cross-scale connection within a CSRB at any scale allows to learn more abundant features from finer to coarser scales or vice versa. Extensive evaluations and comparisons with existing methods verify the advantages of the proposed CSRnet. In future developments, we will extend the CSRnet to handle more general restoration tasks such as image deblurring and blind deconvolution.

\ifCLASSOPTIONcaptionsoff
  \newpage
\fi



%

%
%
\label{sec:reference}
\bibliographystyle{IEEEtran}
\bibliography{duxiaoting}

%








\end{document}